\documentstyle[12pt]{article}
\def\beq{\begin{equation}}
\def\eeq{\end{equation}}
\def\p{{\rm I\!P}}

\begin{document}

\title{{\hfill \small Extended version} \\ {\hfill \small
       Preprint ITP-95-8E} \\ {\hfill \small gr-qc/9503005} \\
\bigskip \bigskip \bigskip
{ON PILOT WAVE QUANTUM COSMOLOGY}}
\author{Yu.~V.~Shtanov \medskip \\
{\small \it Bogolyubov Institute for Theoretical Physics,
252143 Kiev, Ukraine}}
\date{\small December 20, 1995}
\maketitle

\begin{abstract}
The de~Broglie--Bohm (pilot wave) formulation
of quantum theory appears to be free from the
conceptual problems specific to quantum
mechanics (problem of measurement) and to
quantum cosmology (problem of time). We discuss
the issue of quantum equilibrium which arises
within its context. We then study the extension
of this formulation to the case of gravity and
demonstrate that the foliation of spacetime by
space-like hypersurfaces obtained during the
solution of the quantum problem turns out to be
distinguished in general. This means that quantum
pilot wave dynamics is not invariant with respect
to arbitrary change of foliation, or, in other
terms, that quantum non-locality takes place.
We also discuss general structure of a realistic
pilot wave theory which could describe the universe.
\end{abstract}

\bigskip

\centerline{PACS numbers: 03.65.Bz, 04.60.-m, 98.80.Hw}
\newpage

\section{Introduction}

 Among the conceptual problems of quantum gravity and quantum cosmology that
have not yet been completely resolved there are those specific to quantum
mechanics itself and those specific to combination of quantum theory
and the theory of gravity. Let us briefly review these problems.

 Wavefunctions which we ascribe to our microscopic
invisible quantum systems (such as elementary particles) are eventually
used to represent probability amplitudes for the macroscopic
observable events, for instance, these may be appearences of spots on a
photographic plate, or tracks in a cloud chamber.
This is clear, but this is not the whole issue.
We also wish to understand the operation of a measuring device itself,
that is, we wish to build its physical model. We take our
device to consist of
invisible particles of the very same nature as those which we call elementary
and which we describe by wavefunctions. We then ascribe a wavefunction
to the collection of such particles and call it the
wavefunction of the measuring device.\footnote{This procedure,
leading to a description of {\em macroscopic} objects by a wavefunction
which obeys Schr\"{o}dinger equation and respects superposition
principle, may well
be not so innocent as it appears to be. In connection with this
see \cite{Leggett}.} After this, however, our description of the measuring
apparatus becomes twofold, since there are two kinds of wavefunctions
related to it, with two different but interrelated meanings.
The first kind was assigned to a certain microscopic system (the one
which the device was constructed to take ``measurements'' over) and,
we repeat, it represents probability amplitudes for certain observable
events which can occur with the device.
The second kind was assigned to the device itself, and it receives a vague
meaning of representing its {\em state}, which in its own turn must
correspond to
observer's sensible impressions of the device. Such a twofold
description of the measuring apparatus is redundant, and this is the
source of difficulties and paradoxes of various kinds, the famous
Schr\"{o}dinger cat paradox being one of them. To achieve a coherent
physical description of reality this situation must be overcome.

 Now to specific problems which arise in attempts to construct a viable
quantum theory of gravity and cosmology.
Thus, a closed gravitating system, such as ``the universe as a whole,'' is
taken to be described by a wavefunction which obeys Wheeler--De~Witt
equation and which is independent of any time parameter. In what manner can
such a wavefunction correspond to the observed universe which evolves in
time? To resolve this problem cosmologists usually tried to reduce the
phenomenon of time to simple correlations between physical quantities.
They had to choose one of the variables to assign to it the meaning of time.
The choice is always made arbitrarily, and even with a specific
choice made the time fails to be a universal concept, at best it can be defined
in a limited region of the superspace. Moreover, the very meaning of
``the wavefunction of the universe'' remains highly unclear.
What kind of measurements does it correspond to?
How do these measurements affect the state of the universe?
These and related conceptual difficulties hamper our
understanding of an important issue of
modern cosmology, namely, the relevance of quantum laws to the properties
of the observable large scale structure of the universe.

 Numerous efforts were aimed at achieving a coherent
description of reality preserving both its well-established quantum character
on a microscopic level and its macroscopic classical behaviour. Among the
most frequently discussed nowadays is the so-called ``relative state''
formulation of Everett \cite{Everett} developed further by De Witt
\cite{De Witt} and known as ``many-worlds,'' or ``many-minds'' theory.
Like many other people, we do not find this
theory natural, as it demands belief in the ``reality'' of multitudes
of unperceived parallel worlds, and relies on the (hypothetical) physical
theory of mind. Recent ``consistent history'' approach
\cite{Griffiths,Omnes,GH} some of its proponents link with this
``many worlds'' picture \cite{GH}, and some with the Copenhagen line
of thought \cite{Omnes} with its difficulty to account for uniqueness
of the observed reality.
Another interesting in this respect proposal is the pilot wave formulation
of Bohm \cite{Bohm} (for a modern reviews see \cite{BH,BHK,Holland})
which is closely related logically to the well-known ideas of de Broglie
\cite{de Broglie}. In this theory (highly advertised by Bell \cite{Bell})
conceptual difficulties of quantum mechanics mentioned above
appear to acquire simple resolution. It is this theory which will be
the matter of our discussion.

 The basic idea of the pilot wave theory is very simple.
Any physical system is described
by a deterministic evolution of configuration variables (which
Bell has called ``{\em be}ables'' \cite{Bell}). These are
the same as in classical physics and are just the spatial coordinates of the
elementary
particles and the field configurations. The only difference between classical
and quantum theory is in the dynamics of these configuration variables.
In classical physics
their dynamics is determined by the principle of extremal action, or by any
of the equivalent principles. In quantum physics the evolution of the
configuration variables is guided (piloted, in de~Broglie's terminology)
by a quantum wave which obeys Schr\"{o}dinger equation.
This formulation of quantum mechanics eventually has been called
``ontological interpretation'' by its author \cite{BH}.
We shall call it pilot wave interpretation.

 Pilot wave interpretation has been already developed for relativistic
theory of particles and bosonic (scalar and vector) fields
\cite{BH,BHK,Holland}.
It was argued to be consistent with
observable special relativity \cite{BH,Valentini}.
The case of quantum field theory
is only sketched in \cite{BH} and still remains to be developed.
A straightforward extension to the case of general relativity was made
in \cite{Holland}
(minisuperspace pilot wave cosmologies were considered in \cite{KVSCW}).
In this paper we perform further study of such an
extension (Section~\ref{gravity}) and discuss possible ``beable''
formulation of a realistic quantum theory (Section~\ref{discussion})
which could describe our universe. Before doing this we recall the
basics of the theory in more detail (the following section) and discuss
the important question of quantum probabilities (Section~\ref{probabilities}).

\section{Pilot wave quantum dynamics}

 A set of $N$ nonrelativistic spinless particles are described by their
spatial coordinates ${\bf x} \equiv \{{\bf x}_1, \, \ldots, \,
{\bf x}_N\}$. The wavefunction $\psi$ of this system
obeys the Schr\"{o}dinger equation
\beq
i \hbar {\partial \psi \over \partial t} = - {\hbar^2 \over 2}
\sum_n {1 \over m_n} \triangle_n \psi + V \psi \, , \label{schrod}
\eeq
where $V = V({\bf x})$ is the particle interaction potential.
If one represents the wavefunction in the polar form as $\psi =
R \exp \left( i S / \hbar \right)$ then from (\ref{schrod}) it follows that
the phase $S({\bf x}, \, t)$ and the amplitude $R({\bf x}, \, t)$ satisfy the
following system
\beq
{\partial S \over \partial t} + \sum_n {1 \over 2m_n}
\left(\nabla_n S \right)^2 + V + Q = 0 \, , \label{qhj}
\eeq
\beq
{\partial R^2 \over \partial t} + \sum_n {1 \over m_n} \nabla_n \left(R^2
\nabla_n S \right) = 0 \, ,
\eeq
where
\beq
Q = - \sum_n {\hbar^2 \over 2m_n} {\triangle_n R \over R} \label{qpot}
\eeq
is the so-called quantum potential.
In the pilot wave interpretation of quantum mechanics the evolution of
the coordinates ${\bf x}$ is
governed by the phase $S({\bf x}, \, t)$ via the guidance equation
\beq
m_n \dot {\bf x}_n = \nabla_n S \, . \label{guidcon}
\eeq
Eq.~(\ref{qhj}) is the quantum generalization of the classical
Hamilton--Jacobi equation and differs from the latter only by the presence
of the quantum potential $Q({\bf x}, \, t)$.
The guidance condition (\ref{guidcon}) is just the same as in the classical
theory.
In the limit in which the quantum potential $Q$ in (\ref{qhj}) can be
neglected we recover classical evolution. We thus see that new formulation
of quantum theory can be regarded as just a ``deformation'' of the classical
dynamics (general discussion of this analogy with the classical case can be
found in \cite{Holland}).

 In the relativistic theory of half-integer spin one continues to
describe particles by the same configuration variables, namely,
their spatial coordinates. The guidance
conditions and the equations for the wave are now different from those of
nonrelativistic case \cite{BH}. For a system of spin-$1/2$
particles these
are the well known Dirac equations for the multispinor
$\psi_{\alpha_1 \ldots \alpha_N} ({\bf x}_1, \, \ldots, \, {\bf x}_N, \, t)$:
\beq
i \dot \psi_{\alpha_1 \ldots \alpha_N} = \sum_n \left(H_D^{(a)} \psi
\right)_{\alpha_1 \ldots \alpha_N} \, , \label{dirac}
\eeq
where $H_D^{(a)}$ is the usual Dirac matrix Hamiltonian
\beq
H_D = - i \gamma^0 \gamma^i \nabla_i + m \gamma^0 \, , \label{dh}
\eeq
which acts on the
spinor index $\alpha_n$ and on the corresponding argument ${\bf x}_n$ of the
multispinor $\psi_{\alpha_1 \ldots \alpha_N}$. The guidance condition is
\beq
{d x^\mu_n \over d t} = {\psi^\dagger \left( \gamma^0
\gamma^\mu \right)_n \psi \over \psi^\dagger \psi} \, , \label{guide}
\eeq
where the label $n$ numerates the arguments of
the multispinor $\psi$ and $\left( \gamma^\mu \right)_n$ act
on the corresponding spinor index $\alpha_n$.
The multispinor $\psi$ must be chosen antisymmetric with respect to
interchange of any pair of its arguments in accordance with Pauli principle.

 For integer spin the formulation in which the role of
configuration variables would be played by particle coordinates appears
to be impossible \cite{BH,BHK,Holland}.
Instead one has to consider the field spatial configurations as
fundamental configuration variables guided by the corresponding wave
functionals. For
example, the wave functional $\chi[\phi({\bf x}), \, t]$ for a scalar field
$\phi$ will obey the standard Schr\"{o}dinger equation (see Eq.~(\ref{last})
below for the case of curved spacetime background) and the guidance equation
will be
\beq
\dot \phi ({\bf x}, \, t) = \left. {\delta \over \delta \phi ({\bf x})} S [\phi
({\bf x}), \, t] \right|_{\phi ({\bf x}) = \phi ({\bf x}, \, t)}
\, , \label{guide-1}
\eeq
where $S[\phi ({\bf x}), \, t]$ is $\hbar$ times the phase of the wave
functional $\chi[\phi({\bf x}), \, t]$.

 The classical limit in the dynamics of a physical system is achieved
for those configuration variables for which quantum
potential becomes negligible. It is straightforward to see that in this
case such variables evolve in time according
to classical laws of motion.
We send the reader to the reviews \cite{BH,BHK,Holland} for details.
Note that in the new interpretation the temporal dynamics of
the particle coordinates and bosonic field configurations completely determine
the state of a physical system,
be it microscopic or macroscopic. The role of wavefunction in all
physical situations is also one
and the same, namely, to provide the guidance laws for configuration
variables. Therefore the description of all the physical systems has now
become unified, and the source of the difficulties, which lied in the twofold
charachter (mentioned in the Introduction) of this description, has been
thereby eliminated.

\section{Quantum probabilities} \label{probabilities}

 The formalism of quantum dynamics outlined above can be readily applied
to the case of a single closed quantum system. In practice however we
usually deal with what we call quantum ensembles which
are collections of identical systems each piloted in a way decribed in
the preceding section. If all these systems are piloted by one and the same
wavefunction then an ensemble is called pure. Otherwise it is called mixed.
In pilot wave formulation of quantum mechanics the measurement process is
regarded as just a partial case of the generic evolution guided by a wave
function which obeys Schr\"{o}dinger equation.
Probabilistic character of the measurement outcome
arises because of our ignorance of and inability to control the actual
(initial) values of particle and field configuration variables in each
system of an ensemble as well as in the measuring apparatus.

 In order that the probabilities of different measurement outcomes
coincide with those calculated in the standard approach it is necessary
to assume that the configuration variables of the systems in a pure quantum
ensemble
are distributed in accord with their wavefunction,
so that $p(x) = |\psi(x)|^2$, where $x$ denotes the
set of all configuration variables, and $p(x)$ is their
distribution function.
Such a condition is sometimes called {\em quantum equilibrium} condition
\cite{Valentini,DGZ}:
it is a consequence of the Schr\"{o}dinger
equation that provided the equality holds initially for a given ensemble,
it will hold at all times (so long as the ensemble remains closed).
However, in the framework of the theory discussed, one has to demonstrate
how such a distribution arises.\footnote{In
the modified pilot wave proposal of Bohm and Vigier \cite{BV}
(see also \cite{BH})
an additional external stochastic force was added to the right-hand
side of the guidance equations (\ref{guidcon}) in order to account for
the occurrence of quantum equilibrium. In this paper we consider only the
original ``minimal'' version of the pilot wave theory as it is expressed
in \cite{Bohm}.}
Since this question seems to be very important for the pilot
wave interpretational scheme, we shall discuss it in this section.

 Among the recent approaches to the problem of quantum equilibrium that
we are aware of, one is due to Valentini \cite{Valentini}
and another is due to D\"{u}rr, Goldstein and Zangh\`{\i} \cite{DGZ}.
To our mind the proofs and demonstrations contained therein are
essentially incomplete, for the reasons that follow.

 Valentini \cite{Valentini} tries to prove that for a pure ensemble
of closed complicated systems the
{\em coarse-grained} distribution $\overline p(x)$ of the configuration
variables will approach the
coarse-grained value $\overline{|\psi(x)|^2}$ (here overline
denotes coarse graining). The corresponding analysis involves
the quantity $\overline S = - \int \overline p \log\left(\overline p /
\overline{|\psi|^2}\right)$ called ``subquantum entropy.'' By analogy with
classical statistical mechanics (Boltzmann's $H$-theorem) it is
suggested that this quantity should increase in time
approaching its maximum value of zero, thereby leading to coarse-grained
quantum equilibrium. Such a ``proof'' inherits the well-known
difficulties of the ``proof'' of Boltzmann's $H$-theorem.
In fact, from time reversibility of the pilot wave formalism
it only follows that if the conditions $p = \overline p$ and $|\psi|^2
= \overline{|\psi|^2}$ (the conditions of ``no fine-grained microstructure,''
assumed by Valentini to hold at the initial moment of time)
are valid then the above-presented ``entropy'' $\overline S$
acquires its local minimum at that moment of time. There is no proof
that this value will approach zero as time goes to infinity.

 Demonstration of D\"{u}rr {\em et~al.\@} \cite{DGZ} is based on the
notion of {\em typicality} which is applied to the domain of all possible
initial conditions of a model universe. Specifically, the modulus squared
$\left|\Psi\right|^2$ of the universal wave
function is taken to represent the measure density of typicality
in the domain of configuration variables. The preference of such a
measure is based on its property of being time-equivariant.
The authors then show that the
set of those initial conditions which conform (to certain precision) with
all usual quantum mechanical statistical predictions has measure of
typicality close to one. To our mind, equivariance alone of the
{\em specific subjective} measure introduced, although very important
property, is not sufficient for regarding this measure as
relevant to {\em objective} distributions encountered in the
experiments.

 We think that {\em ergodicity} or {\em mixing} argument of some kind
is also required to justify the quantum equilibrium hypothesis.
(In the paper of D\"{u}rr {\em et al.\@} \cite[section 13]{DGZ}
this kind of argument was announced to be of no necessity for reason
which is not very clear to us.)
To explain this, compare the problem of quantum
equilibrium with the problem of classical statistical equilibrium.
In classical statistical mechanics it is the (usually hypothetical)
{\em ergodicity} property of the Hamiltonian flow that can distinguish
the invariant Liouville measure.
Indeed, as a consequence of Birkhoff--Khinchin ergodic theorem,\footnote{For
an introduction to ergodic theory see \cite{HWS}.} the average
time spent by an ergodic system in a region $\Omega$ of its dynamical
variables tends to a value proportional to the invariant measure of this
region as time goes to infinity. In the case of Hamiltonian dynamics such
an invariant measure is the Liouville measure. Note, that the ergodicity
property can be
formulated in terms of {\em any} measure equivalent\footnote{Two measures,
$\mu$ and $\nu$, with common domain are said to be equivalent if $\mu (A) = 0
\Leftrightarrow \nu (A) = 0$ for any set $A$.} to the invariant measure,
in this sense ergodicity
does not rely on this latter. Justification of microcanonical equilibrium
distribution then reduces to the proof (which is usually a non-trivial
task) or assumption of ergodicity property of a particular system.
All this appears
to be of similar relevance to the case of pilot wave quantum mechanics,
so that to establish equilibrium property of measure based on
$\left| \Psi \right|^2$ one not only has to distinguish this measure on
the grounds of its equivariance, but also must relate it to ergodicity of
some kind.

 The fact that the measure with density $\left| \Psi \right|^2$ is only
equivariant rather than also invariant might call for essential
modification of the above argument as compared to the classical case.
However, if we proceed to the {\em universal} level (as suggested in
\cite{DGZ}) and take into account {\em general covariance} of the
complete theory which includes gravity we will find out that the
universal wavefunction does not depend on time\footnote{In some
non-standard proposals, like, e.g., in \cite[b]{Valentini} the
universal wavefunction {\em does} depend on time and does not respect the
Wheeler--De~Witt constraint. To such proposals our argument will not
be applicable.} (which is a well-known fact, see also
Section~\ref{gravity} below) so that the
corresponding ``measure density'' {\em is} invariant. Instead, here arises
the difficulty that the configuration space of the full theory is
infinite-dimensional. We could start surmounting this difficulty by
approximating our ``universe'' by a system with finite number of
degrees of freedom (minisuperspace approach), each time with zero
total energy, and applying ergodicity argument. Of course, there are
states which do not lead to ergodic evolution (like, e.g., a state
with a real wavefunction). One would have to {\em assume} that our
universe (or maybe its part which is of relevance) is rather in a
state which is close to ergodic. Another serious difficulty is that
even in minisuperspace approach the universal wavefunction is usually
not square-integrable, hence, cannot define a finite measure.
To overcome this we can assume that a large subsystem of the universe is
essentially closed and disentangled from the rest of the world,
and that its own wavefunction (which necessarily also will not depend
on time) is square-integrable (at least on a minisuperspace level).
We shall keep in mind these ideas in principle but shall not
develop them here in further detail.

 Suppose that the above difficulties have been somehow overcome
(or did not arise at all, for example, consider a large closed
system in Minkowski background non-dynamical spacetime) and that we
are left with a system in a stationary state with square-integrable
wavefunction. The ergodicity argument could then run as follows.
Let $z = (x, \, y)$ denote the universal configuration variables,
where $x$ represents the coordinates of the system of interest, and
$y$ the coordinates of the environment (compare with the discussion in
\cite{DGZ}). Let the universal wavefunction $\Psi (z)$ have a structure
\beq
\Psi (z) = \psi(x) \phi (y) + \Psi_0 (z) \, , \label{psiuni}
\eeq
for which $\phi (y)$ is non-vanishing in a region $\Omega$, which is
complementary to the $y$-support of $\Psi_0 (z)$.
Then each time the corresponding piloted configuration
variable $Y$ gets into the region $\Omega$, the configuration
variable $X$ is piloted by the wavefunction $\psi (x)$. For $Y$ in
$\Omega$, the probability that $X$ will be in a region $\omega$ will be
given by the limit of the corresponding mean time ratio as follows
\beq
\p(X \in \omega \, | \, Y \in \Omega) = \lim_{T \to \infty} {\int_0^T
f_{\omega \times \Omega} \left( Z (t) \right)\, d t \over \int_0^T
f_{D \times \Omega} \left( Z(t) \right)\, d t } \, , \label{p}
\eeq
where $D$ is the whole domain of $x$, and $f_M$
denotes the characteristic function of the set $M$. If the evolution
is ergodic, then, according to Birkhoff--Khinchin ergodic theorem
(see \cite{HWS}), the limit in (\ref{p}) exists for almost every
initial value of $Z$, and the probability (\ref{p}) is equal to
\beq
\p(X \in \omega \, | \, Y \in \Omega) = {\mu_\Psi \left(\omega \times
\Omega \right) \over \mu_\Psi \left( D \times \Omega \right)}
= \mu_\psi (\omega) \equiv \int \limits_\omega \left| \psi (x)
\right|^2 d x \, , \label{p1}
\eeq
where $\mu_\Psi$ and $\mu_\psi$ are the measures in the
domains, respectively,
of $z$, and of $x$ with densities determined by the corresponding
wavefunctions, and we took the wavefunction $\psi (x)$ to be normalized.
The characteristics of the region $\Omega$ dissappear
from the result (\ref{p1}), and we can apply a formal limit of
infinite-dimensional domain of $y$. To a certain extent the equality
(\ref{p1}) constitutes the proof of quantum equilibrium.

\begin{figure}
 \begin{picture}(400,200)
 \begin{thicklines}
  \put(150,10){\line(0,1){85}}
  \put(150,105){\line(0,1){85}}
  \put(153,10){\line(0,1){85}}
  \put(153,105){\line(0,1){85}}
  \put(250,10){\line(0,1){50}}
  \put(250,70){\line(0,1){60}}
  \put(250,140){\line(0,1){50}}
  \put(247,10){\line(0,1){50}}
  \put(247,70){\line(0,1){60}}
  \put(247,140){\line(0,1){50}}
  \put(350,10){\line(0,1){180}}
  \put(347,10){\line(0,1){180}}
  \put(150,10){\line(1,0){3}}
  \put(150,95){\line(1,0){3}}
  \put(150,105){\line(1,0){3}}
  \put(150,190){\line(1,0){3}}
  \put(247,10){\line(1,0){3}}
  \put(247,60){\line(1,0){3}}
  \put(247,70){\line(1,0){3}}
  \put(247,130){\line(1,0){3}}
  \put(247,140){\line(1,0){3}}
  \put(247,190){\line(1,0){3}}
  \put(347,10){\line(1,0){3}}
  \put(347,190){\line(1,0){3}}
 \end{thicklines}
  \put(135,108){A}
  \put(255,73){B}
  \put(255,143){C}
  \put(355,18){D}
 \end{picture}
 \caption{Two-slit experiment} \label{fig1}
\end{figure}
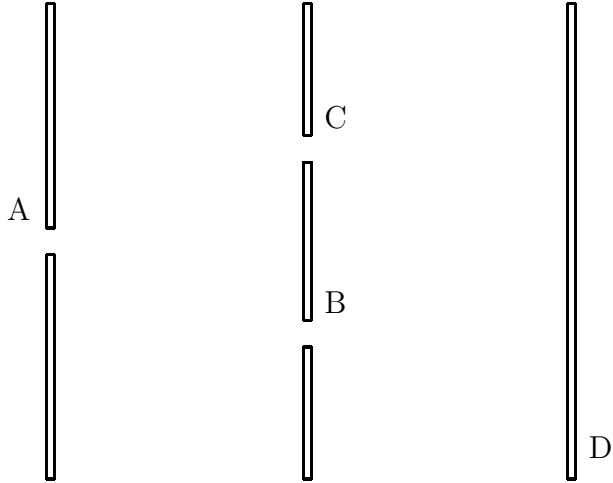

 Certain cases, e.g., measurements of coordinates, appear to be tractable
in a rather simple way. Consider, for example,
the classical two-slit experiment (see Fig.~\ref{fig1}). A system of
collimating instruments and velocity selectors (not shown in the figure)
to the left of the slit $A$
filters out particle wavefunctions, letting those with wavelengths in a
sufficiently narrow band to pass to reach the slit $A$. The role of the slit
$A$ is to produce spherical monochromatic waves in the space to the right
of $A$, for this its dimensions have to be much smaller than the wavelength
of the wavefunction in the space to the left of $A$. The spherical waves
produced are then diffracted
on a pair of slits $B$ and $C$. Now, the appearence of the familiar
interference pattern on the screen $D$ will take place provided particles fall
onto the slit $A$ uniformly within the slit's range (since the
wavefunction is also uniform on the scale of the slit dimensions, the condition
$p = |\psi|^2$ will hold in the vicinity of the slit $A$, hence, by
equivariance argument it will hold also in the space to the right of $A$,
in particular,
on the screen $D$). But this last condition can be easily granted since the
dimensions of the slit $A$ are small. Thus, whatever of {\em continuous}
particle distributions is realized to the left of $A$ (its spatial scale
of continuity is comparable to the spatial scale of particle wavefunctions),
the familiar interference pattern will appear on the screen.

 To what extent such kind of argument can be applied
to other quantum experiments such as, for example,
Stern--Gerlach experiment, remains to be an open question which
we do not attempt to discuss in this paper.

We would only like to point out that sometimes the strong condition of
quantum equilibrium achieved on a {\em universal} level
is not necessary. And, to support the
agreement between the real quantum experiments and the predictions of the
pilot wave theory it is only necessary that quantum equilibrium distribution
arises in {\em preparation} processes (natural or artificial).
For example, in the case of the two-slit experiment (Fig.~\ref{fig1})
quantum equilibrium distribution arises in the space to the right of the
slit $A$, even though it may not take place to the left of $A$.

 If the probabilities as predicted by pilot wave theory turn out to be
the same as in the ordinary quantum theory, then the
problem of relativistic invariance of the pilot wave dynamics will also be
removed.
The problem arises because quantum dynamics of an individual system does
not respect relativistic invariance of the corresponding classical theory:
there is a distinguished reference frame in which the guidance
condition (\ref{guide}) or (\ref{guide-1}) holds. This feature must be
called quantum non-locality. Below we show that the same feature takes place
also for the case of general relativity. Nevertheless, as far as statistical
predictions of the theory are concerned (which are the only testable
predictions), these ({\em if}) coinciding in the ordinary and in the pilot
wave formulations of the quantum theory turn out to be in accord with the
corresponding symmetries (see \cite{BH,Holland} for discussion).

\section{Pilot wave quantum gravity} \label{gravity}

 We now proceed to the second topic of this work, namely, to the question of
pilot wave quantum gravity and quantum cosmology.
The theory of gravity being a particular case of the theory of bosonic
fields, it is readily described in terms of the general formalism outlined
above with modifications caused by the presence of constraints in the theory.
The classical action for a general relativistic system of bosonic
fields in the ADM form looks like follows
\beq
I = \int_M d^3 {\bf x} \, dt \left( \pi^{ab} \dot g_{ab} + \pi_\Phi \dot
\Phi - {\cal N} {\cal H} - {\cal N}^a {\cal H}_a \right) \, , \label{action}
\eeq
where ${\cal H}$ and ${\cal H}_a$ are correspondingly the
so-called Hamiltonian and
momentum constraints, and ${\cal N}$ and ${\cal N}^a$ are the Lagrange
multipliers.  The symbol $\Phi$ denotes the set of bosonic fields, $g_{ab}$
is a positive definite three-metric with the determinant $g$, $\pi_\Phi$ and
$\pi^{ab}$ are the
corresponding generalized momenta. General relativistic constraints have the
following form\footnote{The symbol ``$\approx$'' in (\ref{constr-1}) and
(\ref{constr-2}) below
does not mean approximation, but indicates that the equality to zero is a
constraint.} (we assume for simplicity that there are no other constraints)

\beq
{\cal H} \equiv {1 \over 2 \mu} {\cal G}_{abcd} \pi^{ab}
\pi^{cd} + \mu \sqrt g \left(2 \Lambda - {}^{(3)} {\cal R} \right) + {\cal
H}^\Phi \approx 0 \, , \label{constr-1}
\eeq
\beq
{\cal H}_a \equiv - 2 \nabla_b \pi^b_a + {\cal H}_a^\Phi \approx 0 \, ,
\label{constr-2}
\eeq
where we have written explicitly only the gravitational parts of the
constraints, ${\cal H}^\Phi$ and ${\cal H}_a^\Phi$ are their
$\Phi$-parts which will not be specified here,
$\mu = \left( 16 \pi G \right)^{- 1}$, $G$ is the Newton's constant,
\beq
{\cal G}_{abcd} = {1 \over \sqrt g}
\left(g_{ac} g_{bd} + g_{ad} g_{bc} - g_{ab} g_{cd} \right) \label{wsm}
\eeq
is Wheeler's supermetric, $\nabla_a$ denotes covariant derivative with
respect to the three-metric $g_{ab}$, ${}^{(3)} {\cal R}$ is the scalar
three-curvature of this metric, and $\Lambda$ is the cosmological
constant. The
classical equations of motion, together with the constraint equations, are
obtained by varying the action (\ref{action}) over all its variables. For
example, equations of motion for the metric are
\beq
\dot g_{ab} = {{\cal N} \over \mu} {\cal G}_{abcd} \pi^{cd} +
\nabla_a {\cal N}_b + \nabla_b {\cal N}_a \, , \label{eqmot}
\eeq
where the overdot denotes the derivative with respect to time $t$.

 Proceeding to quantization, first, we recall that in Schr\"{o}dinger
representation
the general relativistic quantum system is described by the wave functional
$\Psi [g_{ab}({\bf x}), \, \Phi({\bf x})]$ where $g_{ab} ({\bf x})$
and $\Phi({\bf x})$ are field configurations on a
three-manifold $\Sigma$ with the coordinates ${\bf x}$. The wave functional
obeys the quantum constraint equations
\beq
\hat {\cal H}_\mu \Psi = 0 \, , \label{qceq}
\eeq
in which $\hat {\cal H}_\mu$ are
obtained from their classical counterparts ${\cal H}_0 \equiv {\cal H}$
and ${\cal H}_a$ after replacement of the
generalized momenta $\pi^{ab}$ and $\pi_\Phi$ by the corresponding
variational derivative operators.\footnote{In this paper we ignore the question
of operator ordering and regularization.}
The wave functional $\Psi$, which is called
the wavefunction of the universe, does not depend on time variable $t$.
This well-known fact, which we mentioned in the beginning of this paper,
remains to be one of the main obstacles in attempts to give meaning
to the wavefunction $\Psi$\@. In the pilot wave interpretation
which we discuss here this problem does not arise.

 To give pilot wave
interpretation of the wavefunction of the universe (see also \cite{Holland})
let us write it in the
standard polar form as $\Psi = R \exp \left( iS / \hbar \right)$ and
substitute into the constraint
equations (\ref{qceq}). The momentum constraints ${\cal H}_a$ will
mean reparametrization invariance of both the amplitude and the phase
of the wave functional. This means that these
functionals can be considered as functionals of three-geometry ${}^{(3)}
{\cal G}$ and of field configuration $\Phi$ regarded in a coordinate-invariant
manner. The Hamiltonian constraint
${\cal H}$ will give birth to two equations which we shall write
in a symbolic form as
\beq
{1 \over 2 \mu} \delta S \circ \delta S + \mu \sqrt g \left(2 \Lambda - {}^
{(3)} {\cal R} \right) - {\hbar^2 \over 2 \mu} {\delta \circ \delta R \over
R} + {\Re \left( \Psi^\dagger \hat {\cal H}^\Phi \Psi\right)
\over R^2} = 0 \, , \label{ehj}
\eeq
\beq
\delta \circ \left( R^2 \delta S \right) - {2 \mu \over \hbar} \Im
\left( \Psi^\dagger
\hat {\cal H}^\Phi \Psi \right) = 0 \, , \label{prob}
\eeq
where $\Re$ and $\Im$ denote real and  imaginary parts respectively, $\delta$
symbolizes
the variational derivative $\delta / \delta g_{ab} ({\bf x})$,
and the circle ``$\circ$''
the contraction with respect to Wheeler's supermetric (\ref{wsm}).
Note that in the formal classical
limit $\hbar \rightarrow 0$ the equation (\ref{ehj}) reduces to the classical
Einstein--Hamilton--Jacobi equation.

 According to the general guidance rules quantum evolution of the
configuration variables
$g_{ab}$ is now given by the equations (\ref{eqmot}) with the substitution
\beq
\pi^{ab} ({\bf x}) \rightarrow \left. {\delta S \over \delta g_{ab} ({\bf x})}
\right|_{g_{ab} ({\bf x}) = g_{ab} ({\bf x}, \, t)} \, . \label{substitute}
\eeq
As for the Lagrange multipliers ${\cal N}$ and ${\cal N}^a$ which enter the
equations (\ref{eqmot}), they remain undetermined and are to be specified
arbitrarily. This situation is analogous to the situation in classical
theory in which the arbitrariness in the choice of the Lagrange
multipliers ${\cal N}$ and ${\cal N}^a$ reflects reparametrization freedom.
Thus to get a solution $g_{ab} ({\bf x}, \, t)$, $\Phi ({\bf x}, \, t)$ to the
quantum dynamics one must first solve the constraint equations (\ref
{qceq}), then specify the initial configurations (say, at $t = 0$) for the
fields $g_{ab}$ and $\Phi$, specify in an arbitrary way the functions
${\cal N} ({\bf x}, \, t)$ and ${\cal N}^a ({\bf x}, \, t)$ and then solve the
guidance equations (\ref{eqmot}) and the analogous equations for $\Phi$.
Solution thus obtained will represent a four-geometry foliated by
spatial hypersurfaces $\Sigma (t)$ on which the three-metric induced is
$g_{ab} ({\bf x}, \, t)$, the lapse function is ${\cal N} ({\bf x}, \, t)$,
the shift vector is ${\cal N}^a ({\bf x}, \, t)$ and the field configuration
is $\Phi ({\bf x}, \, t)$.

 In the classical theory the choice of the Lagrange multipliers
${\cal N}$ and ${\cal N}^a$
affects not the physical solution but only the way in which it is foliated
by a family of hypersurfaces $\Sigma(t)$ and the way in which coordinates
are chosen on each of these hypersurfaces.
If we examine the guidance equations (\ref{eqmot}), (\ref{substitute})
we will notice that the role of the shift vector ${\cal N}^a$
in quantum dynamics is actually analogous to that of the classical theory.
Namely, its form is related only to the way in which the spatial
coordinates ${\bf x}$ are chosen on each of the hypersurfaces $\Sigma(t)$.
This follows
from the fact that the wave functional $\Psi$ is reparametrization invariant.
However the role of the lapse function ${\cal N}$ in the classical and
in the quantum cases is different. Because of nonlocal character
of the equation (\ref{ehj}), which involves variational derivative of a
nonlocal functional $R[g_{ab} ({\bf x}), \, \Phi ({\bf x})]$, physical
character of a solution (four-geometry, for example) {\em will} in
general depend on the specification of the Lagrange
multiplier ${\cal N}$, regarded as a three-scalar on every hypersurface
$\Sigma(t)$. This dependence will be negligible, however, in the
classical limit when the quantum potential in (\ref{ehj}) can be neglected.

 This fact, that the solution to the quantum dynamics depends on the
specification of
the lapse function, signifies that quantum dynamics breaks
foliation-invariance of the classical theory, or, in other terms, that
quantum non-locality takes place. We shall clarify this point by
posing the following question. Suppose that the four-geometry ${}^{(4)}
{\cal G}$ arises as a solution
\beq
{}^{(4)} {\cal G} \equiv \left\{ g_{ab} ({\bf x}, \, t), \;
{\cal N}^\mu ({\bf x}, \, t) \right\} \, , \label{old}
\eeq
to the quantum pilot wave geometrodynamics of pure gravity
described above. If we change in an arbitrary way the space-like foliation
of the four-geometry
obtained ${}^{(4)} {\cal G}$ we will obtain different representation
\beq
{}^{(4)} {\cal G} \equiv \left\{\widetilde g_{ab} ({\bf x},\, t), \;
\widetilde  {\cal N}^\mu ({\bf x}, \, t) \right\} \, ,
                    \label{new}
\eeq
of the {\em same} four-geometry. The question is whether this
new representation is a solution of the quantum pilot wave geometrodynamics
(perhaps with a new solution for the wave functional). The answer is:
generally speaking, no, and this answer distinguishes the space-like
foliation (\ref{old}) from a generic one (\ref{new}).
A simple argument can be given to justify our statement.
Although the classical Hamiltonian constraint equation (\ref{constr-1})
is not satisfied
for the four-geometry (\ref{old}), the classical momentum constraint
equation (\ref{constr-2}) {\em is} satisfied.
Then, however, because the four-geometry (\ref{old}) is not a solution to
classical field equations, if we change foliation and proceed to
(\ref{new}) we must expect the classical momentum constraint equation
(\ref{constr-2}) to be no longer valid. Hence there is no possibility
to obtain (\ref{new}) as a solution to the quantum dynamics.

 To illustrate this conclusion consider a real solution $\Psi$ to the
Wheeler--De~Witt equation, and pick a solution for configuration variables
with arbitrary
initial three-geometry ${}^{(3)} {\cal G}$, arbitrary lapse ${\cal N}$,
and shift ${\cal N}^a \equiv 0$. Since the wave functional is real, the
three geometry will remain constant in time. Now consider an arbitrary
family of space-time foliations $\{\Sigma_\lambda (t)\}$ parametrized by
$\lambda$ from a real interval containing zero, such that $\Sigma_0 (t)$
coincides with the original foliation.
For each foliation from the family $\{\Sigma_\lambda\}$ there
will be induced ${}^{(3)} {\cal G}_\lambda$, ${\cal N}_\lambda$ and
${\cal N}^a_\lambda$ in spacetime. To simplify situation we can
choose spatial coordinates on the hypersurfaces of each
family in such a way that ${\cal N}^a_\lambda \equiv 0$
(this is always possible).
Now, the value of the momentum constraint ${\cal H}_a (\lambda)$
with respect to the
foliation $\Sigma_\lambda$ will not necessarily be zero, in fact one can
easily calculate its derivative at $\lambda = 0$ to be
\beq
\left. {\partial {\cal H}_a (\lambda) \over \partial \lambda}
\right|_{\lambda = 0}
= 2 \mu \sqrt g \left( {\cal N} {}^{(3)} {\cal R}^b_a \nabla_b \xi -
\nabla_a \nabla^b {\cal N} \nabla_b \xi + {}^{(3)} \triangle {\cal N}
\nabla_a \xi \right) \, , \label{deriv}
\eeq
where $\xi$ is the temporal component of the vector field $\partial /
\partial \lambda$ at $\lambda = 0$ generated by the family of foliations
$\{\Sigma_\lambda\}$.
Obviously, the expression (\ref{deriv}) will in general
be non-zero. Therefore change of foliation will lead to a representation
(\ref{new}) which cannot be a solution to quantum dynamics.

 For an example in a finite form consider again a real solution $\Psi$
for the wave functional. Choose in an
arbitrary way the initial three-geometry ${}^{(3)} {\cal G} (0)$,
and choose ${\cal N}^a \equiv 0$ for all $t$, and
${\cal N} \equiv 1$ for $0 \leq t \leq t_1$, ${\cal N} \not \equiv const$
for $t > t_1$. Again, since the wave functional is real the solution will be
${}^{(3)} {\cal G} (t) \equiv {}^{(3)} {\cal G} (0)$. Now change the
foliation in such a way that $\widetilde {\cal N}^a \equiv 0$ for
all $t$,
$\widetilde {\cal N} (t) \equiv {\cal N} (t_1 + t)$ for some interval
$0 \leq t \leq t_2 < t_1$, and $\widetilde {\cal N} \equiv {\cal N}$
for $t > t_1$ (this is always possible in general). The new three-geometry
${}^{(3)} \widetilde {\cal G} (t)$ will be different from the old one
in the interval $0 \leq t \leq t_2$ but will remain unchanged in the
region $t \geq t_1$. We will have then
${}^{(3)} \widetilde {\cal G} (0) = {}^{(3)} \widetilde {\cal G} (t_1)$
(identical initial conditions for time intervals $t \geq 0$ and $t \geq t_1$),
$\widetilde {\cal N}^\mu (t) \equiv \widetilde {\cal N}^\mu (t_1 + t)$
in a certain interval of
$t \geq 0$ (identical lapse and shift in the corresponding time intervals),
but ${}^{(3)} \widetilde {\cal G} (t) \not \equiv {}^{(3)} \widetilde
{\cal G} (t_1 + t)$ (different solutions in the corresponding time intervals)
which shows that new foliation cannot in principle be a solution of quantum
dynamics.

 Let us now consider the standard treatment (see, e.g., \cite{Padmanabhan}
and references therein) of  the classical limit for gravitation and assume the
wavefunction to acquire an approximate shape
\beq
\Psi[{}^{(3)} {\cal G}, \, \Phi] \approx R[{}^{(3)} {\cal G}] \exp
\left( {iS[{}^{(3)}
{\cal G}] \over \hbar} \right) \chi[{}^{(3)} {\cal G}, \, \Phi] \, ,
\label{psi}
\eeq
in which real $S[{}^{(3)} {\cal G}]$ is the solution of the classical purely
gravitational
Einstein--Hamilton--Jacobi equation (which is just the equation (\ref{ehj})
without
the quantum potential for gravity and without the $\Phi$-part),
and real $R[{}^{(3)} {\cal G}]$ is chosen so that it obeys the
equation
\beq
\delta \circ \left( R^2 \delta S  \right) = 0 \, ,
\eeq
the notation of which was explained above. If we assume the dependence of the
functional $\chi[{}^{(3)} {\cal G}, \Phi]$ on the three-geometry to be
sufficiently weak, then from the equations
(\ref{ehj}) and (\ref{prob}) with the quantum potential for gravity
neglected we will obtain the following equation for this functional
\beq
{i \hbar \over \mu} \delta S \circ \delta \chi = \hat {\cal H}^\Phi
\chi \, . \label{last-1}
\eeq
In the classical limit the role of the wave functional $\chi$ for gravity
is negligible. The guidance equations (\ref{eqmot}), (\ref{substitute})
for the metric will then determine the
four-geometry which will obey the classical Einstein equations (because
the functional $S[{}^{(3)} {\cal G}]$ in (\ref{psi}) obeys the classical
Eistein--Hamilton--Jacobi equation). And the functional $\chi[t,
\Phi] \equiv \chi[{}^{(3)} {\cal G} (t), \Phi]$ will
evolve on this background four-geometry according to the standard
Schr\"{o}dinger equation
\beq
i \hbar \dot \chi = \int_\Sigma d^3 {\bf x} \, {\cal N}^\mu \hat
{\cal H}^\Phi_\mu \, \chi \, , \label{last}
\eeq
which follows from (\ref{last-1}).
We thus see how naturally the limit of quantum field theory on a classical
geometric background is attained in the pilot wave formulation.

 To end this section we note that in the pilot wave framework
considered here one can consistently formulate a hybrid theory in which only
part of the degrees of freedom are quantized, and other remain classical.
To this end it is only sufficient to drop the quantum potential term
in Eq.~(\ref{ehj}) for those degrees of freedom which are to be classical
and leave Eq.~(\ref{prob}) unchanged.
For example, if one does not find it appropriate to quantize gravity one
simply should drop the explicit quantum potential term for gravity in
Eq.~(\ref{ehj}).

\section{Extension to a complete theory and discussion} \label{discussion}

 A few points must be taken about the extension of the proposal considered
to the case of full quantum theory (see also some reasoning in
\cite[Section~12.3]{BH}). In constructing such a theory one has first to
resolve the problem
of fermionic negative energy levels. A straightforward way of doing this
will be to assume, following Dirac, that all such energy levels are occupied.
This requires infinite number of particles, hence, infinite number of
arguments
of the wavefunction $\Psi$. This number will be countable if we also assume
the spatial section $\Sigma$ to be compact, for example, we could take it to
be topologically a
three-sphere. After this we must learn to describe interactions between
particles
and bosonic fields. Presumably this will require considering wave
functionals of type
$\Psi_{\alpha_1 \ldots \alpha_n \ldots} [{}^{(3)}{\cal G}, \, \Phi, \,
{\bf x}_1, \, \ldots, \, {\bf x}_n, \, \ldots ]$ which are
simultaneously antisymmetric
multispinors with a countable set of spinor arguments. Gravity will now have
to be described in terms of variables adjusted for coupling to spinors, for
example, in terms of Ashtekar variables (see \cite{Ashtekar}).
The wavefunction would obey constraint equation of the following type
\beq
\left( {\cal H} + \sum_n {\cal H}_D^{(n)} + \sum_n {\cal H}_{\rm int}^{(n)}
\right) \Psi = 0 \, ,
\eeq
where ${\cal H}$ is the Hamiltonian constraint of the bosonic fields,
which acts on the field arguments of the wavefunction,
${\cal H}_D^{(n)}$ is the Dirac Hamiltonian constraint, and
${\cal H}_{\rm int}^{(n)}$ is the interaction Hamiltonian constraint
which act on the $n$-th fermionic particle argument. Interactions will make
possible particle transitions from the negative energy levels to the
positive ones and backwards, which processes will correspond to creation
and annihilation of pairs. Such a theory still remains to be elaborated.

 We end this paper by a few more remarks. First, it is
clear (see also the general discussion in \cite{BH}) that
the wavefunction of the universe need not be a member of a Hilbert space,
that is, it need not be square-integrable with respect to {\em all}
its arguments.
The role of the wavefunction in the pilot wave interpretation is to
provide guidance for the particle and field configuration variables
and this clearly does
not require normalization. Probabilistic nature arises here as a secondary
concept and requires at most only square-integrability over
part of its arguments. This
fact in a simple way removes the second difficulty (besides the
problem of time) of the standard quantum gravity, namely, lack of its
probabilistic interpretation. However, this same fact
creates obstacles on the way of justifying the quantum equilibrium
hypothesis (see Section~\ref{probabilities}).

 Our second remark concerns the choice of boundary
conditions for the wavefunction of the universe. In the pioneer formulation
\cite{HH}
Hartle and Hawking considered their no-boundary conditions as leading
to the real wavefunction (their treatment was afterwards more correctly
specified \cite{Halliwell} and shown to produce besides real also complex
solutions). From
the viewpoint of the pilot wave interpretation a real wavefunction
of the universe is implausible as it leads to the evolution which is hardly
compatible with the observed one. It seems that in the approach considered
the wavefunction with tunneling boundary conditions \cite{Linde,Vilenkin}
is a good candidate for describing the real universe, as it is essentially
complex \cite{Vilenkin} and can attain classical behaviour in certain
regions of superspace (see very similar remarks in \cite{AM}).

 Finally, it is not unlikely that pilot wave interpretation of
quantum theory will help deeper understanding and resolving the famous
information loss paradox \cite{Hawking} as well as other problems of the same
nature which might involve the effects of quantum gravity (see
in this respect \cite{AM}).


\begin{thebibliography}{99}
\bibitem{Leggett} A.~J.~Leggett, Supplement of Progr. Theor. Phys.
    No 69, 80 (1980); \\
    A.~J.~Leggett and A.~Garg, Phys. Rev. Lett. {\bf 54} 857 (1985).
\bibitem{Everett} H.~Everett, Rev. Mod. Phys. {\bf 29}, 454 (1957).
\bibitem{De Witt} B.~S.~De Witt and N.~Graham, {\it The Many-Worlds
   Interpretation of Quantum Mechanics}, (Princeton University Press,
   Princeton, New Jersey, 1973).
\bibitem{Griffiths} R.~B.~Griffiths, J. Stat. Phys. {\bf 36}, 219 (1984); \
   Am. J. Phys. {\bf 55}, 11 (1987); \  Phys. Rev. Lett. {\bf 70}, 2201 (1993).
\bibitem{Omnes} R.~Omn\`{e}s, J. Stat. Phys. {\bf 53}, 893, 933, 957 (1988); \
   {\it ibid.} {\bf 57}, 357 (1989); \ Rev. Mod. Phys. {\bf 64}, 339 (1992); \\
   R.~Omn\`{e}s, {\it The Interpretation of Quantum Mechanics}, (Princeton
   University Press, Princeton, New Jersey, 1994).
\bibitem{GH} M.~Gell-Mann and J.~B.~Hartle, ``Quantum Mechanics in the Light
   of Quantum Cosmology,'' in {\it Proc. 3rd Int. Symp. Found. of Quantum
   Mechanics}, ed.\@ by S.~Kobayashi, (Physical Society of Japan, Tokyo, 1989);
   in {\it Complexity, Entropy, and the Physics of Information}, ed.\@ by
   W.~Zurek (Addison-Wesley, Reading, 1990).
\bibitem{Bohm} D.~Bohm, Phys. Rev. {\bf 85}, 166 (1952); {\it ibid.}
   {\bf 85}, 180 (1952).
\bibitem{BH}  D.~Bohm and B.~J.~Hiley, {\it The Undivided Universe: An
   Ontological Interpretation of Quantum Theory}, (Routledge,
   London and New York, 1993).
\bibitem{BHK} D.~Bohm and B.~J.~Hiley, Phys. Rep. {\bf 144}, 323 (1987);
   \\ D.~Bohm, B.~J.~Hiley and P.~N.~Kaloyerou, {\it ibid.} {\bf 144}, 349
   (1987); \\
   P.~N.~Kaloyerou, {\it ibid.} {\bf 244}, 287 (1994).
\bibitem{Holland} P.~Holland, Phys. Rep. {\bf 224}, 95 (1993); \\
   P.~Holland, {\it The Quantum Theory of Motion},
   (Cambridge University Press, Cambridge 1993).
\bibitem{de Broglie} L.~de~Broglie, J.~Physique, 6e s\'{e}rie {\bf 8}, 225
   (1927); \\ L.~de~Broglie, {\it Une Interpr\'{e}tation Causale et non
   Lin\'{e}aire de la M\'{e}canique Ondulatoire: la Th\'{e}orie de la Double
   Solution}, (Gauthier-Villars, Paris 1956). English translation:
   Elsevier, Amsterdam 1960.
\bibitem{Bell} J.~S.~Bell, {\it Speakable and Unspeakable in Quantum
               Mechanics}, (Cambridge University Press, Cambridge, 1987).
\bibitem{Valentini} A.~Valentini, a)~Phys. Lett. A {\bf 156}, 5 (1991); \
   {\it ibid}. {\bf 158}, 1 (1991). \ \ b)~``On the Pilot-Wave Theory of
   Classical, Quantum and Subquantum Physics,'' Ph.D. Thesis, (ISAS,
   Trieste 1992).
\bibitem{KVSCW} J.~Kowalski-Glikman and J.~C.~Vink, Class. Q. Grav. {\bf 7},
   901 (1990); \\ J.~C.~Vink, Nucl. Phys. {\bf B369}, 707 (1992); \\
   E.~Squires, Phys. Lett. A {\bf 162}, 35 (1992); \\
   C.~Callender and R.~Weingard, Phil. Sci. Ass. {\bf 1}, 218 (1994).
\bibitem{BV} D.~Bohm and J.-P.~Vigier, Phys. Rev. {\bf 96}, 208 (1954).
\bibitem{DGZ} D.~D\"{u}rr, S.~Goldstein and N.~Zangh\`{\i}, J. Stat. Phys.
   {\bf 67}, 843 (1992).
\bibitem{HWS} P.~R.~Halmos, {\it Lectures on Ergodic Theory}, (The
   Mathematical Society of Japan, Tokyo, 1956); \\
   P.~Walters, {\it An Introduction to Ergodic Theory},
   (Springer-Verlag, 1981); \\
   Ya.~G.~Sinai, {\it Topics in Ergodic Theory}, (Princeton
   University Press, Princeton, New Jersey, 1994).
\bibitem{Padmanabhan} T.~Padmanabhan, ``Time, Mach's Principle and Quantum
   Cosmology,'' in {\it Physical Origin of Time Asymmetry}, ed.\@ by
   J.~J.~Halliwell, J.~P\'{e}rez-Mercader and W.~H.~Zurek (Cambridge
   University Press, Cambridge, 1994).
\bibitem{Ashtekar} A.~Ashtekar, {\it Non-perturbative Canonical Gravity},
   (World Scientific, Singapore, 1991).
\bibitem{HH} J.~B.~Hartle and S.~W.~Hawking, Phys. Rev. {\bf D28}, 2960 (1983).
\bibitem{Halliwell} J.~J.~Halliwell and J.~B.~Hartle, Phys. Rev. {\bf D41},
   1815 (1990).
\bibitem{Linde} A.~D.~Linde, Zh. Exp. Teor. Fiz. {\bf 87}, 369 (1984)
   [Sov.Phys. JETP {\bf 60}, 211 (1984)]; \
   Lett. Nuovo Cim. {\bf 39}, 401 (1984).
\bibitem{Vilenkin} A.~Vilenkin, Phys. Rev. {\bf D30}, 509 (1984); \
   {\it ibid.} {\bf D33}, 3560 (1986); \  {\it ibid.} {\bf D37}, 888 (1988).
\bibitem{AM} S.~P.~de~Alwis and D.~A.~MacIntire, Phys. Rev. {\bf D50}, 5164
   (1994).
\bibitem{Hawking} S.~W.~Hawking, Phys. Rev. {\bf D14}, 2460 (1976).
\end{thebibliography}
\end{document}